\documentstyle[amssymb,amstex]{article}

\catcode`\@=11
\def\numberbysection{\@addtoreset{equation}{section}
        \def\theequation{\thesection.\arabic{equation}}}
\numberbysection

\begin{document}

\newlength{\lno} \lno1.75cm \newlength{\len} \len=\textwidth%
\addtolength{\len}{-\lno}

\setcounter{page}{0}

\baselineskip8mm \renewcommand{\thefootnote}{\fnsymbol{footnote}} \newpage %
\setcounter{page}{0}

\begin{titlepage}     
\vspace{0.5cm}
\begin{center}
{\Large\bf On the ${\cal{U}}_{q}[sl(2)]$ Temperley-Lieb reflection matrices}\\
\vspace{1cm}
{\large \bf  A. Lima-Santos } \\
\vspace{1cm}
{\large \em Universidade Federal de S\~ao Carlos, Departamento de F\'{\i}sica \\
Caixa Postal 676, CEP 13569-905~~S\~ao Carlos, Brasil}\\
\end{center}
\vspace{1.2cm}

\begin{abstract}
This work concerns the boundary integrability of the spin-s  ${\cal{U}}_{q}[sl(2)]$ Temperley-Lieb model. 
A systematic computation method is used to constructed the solutions of the boundary Yang-Baxter equations.  
For $s$ half-integer, a general $2s(s+1)+3/2$  free parameter solution is presented. It turns that for $s$ integer, 
the general solution has $2s(s+1)+1$ free parameters. Moreover, some particular solutions are discussed.
\end{abstract}
\vspace{2cm}
\begin{center}
Keywords:  integrable spin chains
(vertex models), solvable lattice models.
\end{center}
\vfill
\begin{center}
\small{\today}
\end{center}
\end{titlepage}

\baselineskip7mm

\newpage

\section{Introduction}

The search for integrable models through solutions of the Yang--Baxter
equation \cite{Baxter, KIB, AAR}%
\begin{equation}
{\cal R}_{12}(u-v){\cal R}_{13}(u){\cal R}_{23}(v)={\cal R}_{23}(v){\cal R}%
_{13}(u){\cal R}_{12}(u-v)  \label{int.1}
\end{equation}%
has been performed by the quantum group approach in \cite{KR}, where the
problem is reduced to a linear one. Indeed, the ${\cal R}$ matrices
corresponding to vector representations of all nonexceptional affine Lie
algebras have been determined in this way by Jimbo \cite{Jimbo}.

A similar approach is desirable for finding solutions of the boundary
Yang--Baxter equation \cite{Chere, Skly} where the boundary weights follow
from $K$ matrices which satisfy a pair of equations, namely the reflection
equation%
\begin{equation}
{\cal R}_{12}(u-v)K_{1}^{-}(u){\cal R}%
_{12}^{t_{1}t_{2}}(u+v)K_{2}^{-}(v)=K_{2}^{-}(v){\cal R}%
_{12}(u+v)K_{1}^{-}(u){\cal R}_{12}^{t_{1}t_{2}}(u-v)  \label{int.2}
\end{equation}%
and the dual reflection equation%
\begin{eqnarray}
&&{\cal R}_{12}(-u+v)\left( K_{1}^{+}\right) ^{t_{1}}(u)M_{1}^{-1}{\cal R}%
_{12}^{t_{1}t_{2}}(-u-v-2\rho )M_{1}\left( K_{2}^{+}\right) ^{t_{2}}(v)= 
\nonumber \\
&&\left( K_{2}^{+}\right) ^{t_{2}}(v)M_{1}{\cal R}_{12}(-u-v-2\rho
)M_{1}^{-1}\left( K_{1}^{+}\right) ^{t_{1}}(u){\cal R}%
_{12}^{t_{1}t_{2}}(-u+v).  \label{int.3}
\end{eqnarray}%
In this case duality supplies a relation between $K^{-}$ and $K^{+}$ \cite%
{MN1}%
\begin{equation}
K^{+}(u)=K^{-}(-u-\rho )^{t}M,\qquad M=V^{t}V  \label{int.4}
\end{equation}%
Here $t$ denotes transposition and $t_{i}$ denotes transposition in the i-th
space. $V$ is the crossing matrix and $\rho $ the crossing parameter, both
being specific to each model \cite{Bazha}.

With this goal in mind, the study of boundary quantum groups was initiated
in \cite{MN2}. These boundary quantum groups have been used to determine $%
A_{1}^{(1)}$ reflection matrices for arbitrary spin \cite{DN}, and the $%
A_{2}^{(2)}$ and some $A_{n}^{(1)}$ reflection matrices were derived again
in \cite{Nepo}. Reflection solutions from ${\cal R}$-matrices corresponding
to vector representations of Yangians and super-Yangians were presented in 
\cite{Arnau}. However, as observed by Nepomechie \cite{Nepo}, an independent
systematic method of constructing the boundary quantum group generators is
not yet available. In contrast to the bulk case \cite{Jimbo}, one cannot
exploit boundary affine Toda field theory, since appropriate classical
integrable boundary conditions are not yet known \cite{BCDR}. Therefore, it
is still an open question whether it is possible to find all solutions of
the reflection equations by using quantum group generators.

Independently, there has been an increasing amount of effort towards the
understanding of two-dimensional integrable theories with boundaries via
solutions of the reflection equation (\ref{int.2}). In field theory,
attention is focused on the boundary $S$ matrix \cite{GZ, FK}. In
statistical mechanics, the emphasis has been laid on deriving all solutions
of (\ref{int.2}) because different $K$-matrices lead to different
universality classes of surface critical behavior \cite{Batch1} and allow
the calculation of various surface critical phenomena, both at and away from
criticality \cite{Batch2}.

Although being a hard problem, the direct computation has been used to
derive the solutions of the boundary Yang--Baxter equation (\ref{int.2}) for
given ${\cal R}$. For instance, we mention the solutions with ${\cal R}$
matrix based in non-exceptional Lie algebras \cite{DeVega, RLS} and
superalgebras \cite{Doikou1, LS}. The regular K-matrices for the exceptional 
${\cal U}_{q}[G_{2}]$ vertex model were obtained in \cite{LSM}$.$ Many
diagonal solutions for face and vertex models associated with affine Lie
algebras were presented in \cite{Batch2}. For {\small A--D--E}
interaction-round face ({\small IRF}) models, diagonal and some non-diagonal
solutions were presented in \cite{BP}. Reflection matrices for
Andrews--Baxter--Forrester models in the {\small RSOS/SOS} representation
were presented in \cite{Ahn}. Apart from these c-number solutions of the
reflection equations there must also exist non trivial solutions that
include boundary degree of freedom as were derived for the sine-Gordon
theory in \cite{BK}.

Here we will again touch this issue in order to include the Temperley-Lieb
lattice models \cite{TL} arising from the quantum group ${\cal U}_{q}[sl(2)]$
\cite{Batch3}.

We have organized this paper as follows. In Section $2$ the model is
presented, in Section $3$ we choose the reflection equations and their
solutions. In Section $4$ some reduced solutions are derived and the Section 
$5$ is reserved for the conclusion.

\section{The model}

The Temperley-Lieb algebra is very useful in the study of two dimensional
lattice statistical mechanics. It provided an algebraic framework for
constructing and analyzing different types of integrable lattice models,
such as $Q$-state Potts model, {\small IRF} model, $O(n)$ loop model,
six-vertex model, etc. \cite{Martin1}.

From the representation of the Temperley-Lieb algebra \cite{Batch3}, one can
build solvable vertex models with the $R$ operator defined by 
\begin{equation}
R(u)=\frac{\sinh (\eta -u)}{\sinh \eta }{\cal I}+\sqrt{Q}\frac{\sinh u}{%
\sinh \eta }{\cal P}_{0}  \label{mod.1}
\end{equation}%
where ${\cal I}$ is the identity operator and ${\cal P}_{0}$ a suitable
projector. Here $u$ is the spectral parameter and the anisotropic parameter $%
\eta $ is chosen so that%
\begin{equation}
2\cosh \eta =\sqrt{Q}  \label{mod.2}
\end{equation}%
For the spin-$s$ ${\cal U}_{q}[sl(2)]$ model, ${\cal I}$ is the $(2s+1)^{2}$
by $(2s+1)^{2}$ unity matrix and $\sqrt{Q}{\cal P}_{0}=$ $U$ \ is the spin
zero operator%
\begin{equation}
<i,j|U|k,l>=(-1)^{i+k}q^{i+k}\delta _{i+j,0}\delta _{k+l,0}\qquad
\{i,j,k,l\}=-s,-s+1,..,s  \label{mod.3}
\end{equation}%
and $\sqrt{Q}=[2s+1]=q^{2s}+q^{2(s-1)}+\cdots +q^{-2s}.$

The Hamiltonian limit 
\begin{equation}
R(u)={\cal I}+u(\alpha ^{-1}{\cal H}+\beta {\cal I})  \label{mod.4}
\end{equation}%
with $\alpha =\sinh \eta $ , $\beta =-\coth \eta $ leads to the quantum spin
chains 
\begin{equation}
{\cal H}=\sum_{k=1}^{N-1}U_{k,k+1}+{\rm bt}  \label{mod.5}
\end{equation}%
where, instead of periodic boundary condition, we are taking into account
the existence of integrable boundary terms ${\rm bt}$ \cite{Skly}, derived
from the $K^{-}$ and $K^{+}$ matrices presented in the next sections.

We also have to consider the permuted operator ${\cal R}=PR$ which is
regular satisfying {\small PT}-symmetry, unitarity and crossing symmetry%
\begin{eqnarray}
{\cal R}_{12}(0) &=&P,  \nonumber \\
{\cal R}_{12}^{t_{1}t_{2}}(u) &=&P{\cal R}_{12}(u)P={\cal R}_{21}(u), 
\nonumber \\
{\cal R}_{12}(u){\cal R}_{12}^{t_{1}t_{2}}(-u) &=&x_{1}(u)x_{1}(-u)I, 
\nonumber \\
{\cal R}_{12}(u) &=&(-1)^{2s}(V\otimes 1){\cal R}_{12}^{t_{2}}(-u-\rho
)(V\otimes 1)^{-1}  \label{mod.6}
\end{eqnarray}%
where $\rho =-\eta $ is the crossing parameter and $V$ is the crossing
matrix, specified by%
\begin{equation}
V_{i,j}=(-1)^{i-1}q^{s+1-i}\delta _{i,2s+2-j}  \label{mod.7}
\end{equation}%
and $P$ is the permutation matrix $<i,j|P|k,l>=\delta _{i,l}\delta _{j,k}.$

\section{The reflection matrices}

\bigskip The reflection equation (\ref{int.2}) where $K_{1}^{-}=K^{-}\otimes
I$, $K_{2}^{-}=I\otimes K^{-}$, ${\cal R}_{12}={\cal R}$ and ${\cal R}%
_{12}^{t_{1}t_{2}}=P{\cal R}P$, $I$ and $K^{-}$ are $(2s+1)$ by $(2s+1)$
matrices satisfying the normal condition $K^{-}(0)=I$. Substituting%
\begin{equation}
K^{-}(u)=\sum_{i,j=1}^{2s+1}k_{i,j}(u)E_{i,j}  \label{lkm.1}
\end{equation}%
where $(E_{i,j})_{k,l}=\delta _{i,k}\delta _{j,l}$ \ are the Weyl matrices
and ${\cal R}(u)=P[x_{1}(u)I+x_{2}(u)U]$ with 
\begin{equation}
x_{1}(u)=\frac{\sinh (\eta -u)}{\sinh \eta },\qquad x_{2}(u)=\frac{\sinh u}{%
\sinh \eta },  \label{lkm.2}
\end{equation}%
into (\ref{int.2}), we will have $(2s+1)^{4}$ functional equations for the $%
k_{i,j}$ elements, many of them not independent equations. In order to solve
these functional equations, we shall proceed as follows. First we consider
the $(i,j)$ component of the matrix equation (\ref{int.2}). By
differentiating it with respect to $v$ and taking $v=0$, we get algebraic
equations involving the single variable $u$ and $(2s+1)^{2}$ parameters%
\begin{equation}
\beta _{i,j}=\frac{dk_{i,j}(v)}{dv}|_{v=0}\text{ },\qquad i,j=1,2,...,2s+1
\label{lkm.3}
\end{equation}

Analyzing the refection equations one can see that they possess a special
structure. Several equations exist involving only two non-diagonal elements.
They can be solved by the relations

\begin{equation}
k_{i,j}(u)=\frac{\beta _{i,j}}{\beta _{1,2s+1}}k_{1,2s+1}(u)\qquad (i\neq
j=\{1,2,...,2s+1\})  \label{lkm.4}
\end{equation}%
We thus left with several equations involving two diagonal elements and $%
k_{1,2s+1}(u)$. Such equations are solved by the relations%
\begin{equation}
k_{i,i}=k_{1,1}(u)+\left( \beta _{i,i}-\beta _{1,1}\right) \frac{%
k_{1,2s+1}(u)}{\beta _{1,2s+1}}\qquad (i=2,3,...,2s+1).  \label{lkm.5}
\end{equation}%
Finally, we can use the equation $(1,2s+1)$ in order to find the element $%
k_{1,1}(u)$:%
\begin{eqnarray}
k_{1,1}(u) &=&\frac{k_{1,2s+1}(u)}{\beta _{1,2s+1}[x_{2}(u)\cosh \eta
+x_{1}(u)]}\left\{ \frac{x_{1}(u)x_{2}^{\prime }(x)-x_{1}^{\prime
}(u)x_{2}(u)}{x_{2}(u)}-\frac{1}{2}\frac{x_{1}(u)}{\beta _{1,2s+1}}%
\sum_{j=2}^{2s}\beta _{1,j}\beta _{j,2s+1}\right.  \nonumber \\
&&\left. -\frac{1}{2}x_{2}(u)\sum_{j=2}^{2s}(\beta _{j,j}-\beta
_{1,1})q^{2j-2s-2}-\frac{1}{2}\left[ x_{1}(u)+q^{2s}x_{2}(u)\right] (\beta
_{2s+1,2s+1}-\beta _{1,1})\right\}  \label{lkm.6}
\end{eqnarray}%
where $x_{i}^{\prime }(u)=dx_{i}(u)/du$. \ After these steps, we can write
all matrix elements in terms of $k_{1,2s+1}(u)$.

Now, substituting these expressions into the remained equations $(i,j)$, we
are left with several constraint equations involving the $\beta _{i,j}$
parameters.

First, we consider the blocks of four equations \cite{ALS} 
\begin{equation}
B[j,2s]=\{(j,2s),(2s,j),(j^{\prime \prime },(2s)^{\prime \prime
}),((2s)^{\prime \prime },j^{\prime \prime })\}  \label{lkm.7}
\end{equation}%
where $i^{\prime \prime }=(2s+1)^{2}+1-i$, in order to fix the $2s$ diagonal
parameters $\beta _{j,j}$\ , $j=2,3,...,2s$ and $\beta _{2s+1,2s+1}$ \ (the
parameter $\beta _{1,1}$ is fixed by the normal condition).

Subsequently, we are embroiled with a too large task which consist into find
some parameters $\beta _{i,j}$ in terms of the free parameters. For the spin-%
$s$ ${\cal U}_{q}[sl(2)]$ Temperley-Lieb model, the number of free
parameters is too large and we need too large computer resources in order to
express the fixed parameters in term of them.

The dual equation (\ref{int.3}) is solved by the $K^{+}$matrices via the
isomorphism (\ref{int.4}) with $\rho =-\eta $ and the matrix $M$ specified by%
\begin{equation}
M_{i,j}=q^{-2(s+1-i)}\delta _{i,j}.
\end{equation}%
Here we note that trace of the matrix $M$ is equal to $2\cosh \eta $.

Now, we explicitly show these computations for the first cases. Firstly, let
us make the choice%
\begin{equation}
k_{1,2s+1}(u)=\beta _{1,2s+1}\frac{x_{2}(u)[x_{2}(u)\cosh \eta +x_{1}(u)]}{%
x_{1}(u)x_{2}^{\prime }(x)-x_{1}^{\prime }(u)x_{2}(u)}=\frac{1}{2}\beta
_{1,2s+1}\sinh (2u)  \label{lkm.8}
\end{equation}%
in order to simplify our presentation.

\subsection{The ${\cal U}_{q}[sl(2)]$\ Temperley-Lieb K matrix for $s=\frac{1%
}{2}$}

For the case $s=\frac{1}{2}$ we have the well-known three free parameter
solution \cite{GZ}:%
\begin{equation}
K^{-}(u)=\left( 
\begin{array}{cc}
k_{1,1}(u) & \frac{1}{2}\beta _{1,2}\sinh (2u) \\ 
&  \\ 
\frac{1}{2}\beta _{2,1}\sinh (2u) & k_{1,1}(u)+\frac{1}{2}(\beta
_{2,2}-\beta _{1,1})\sinh (2u)%
\end{array}%
\right)  \label{lkm.9}
\end{equation}%
Using (\ref{lkm.8}), the expression (\ref{lkm.6}) has a simplified form 
\begin{equation}
k_{1,1}(u)=1-\frac{1}{2}(\beta _{2,2}-\beta _{1,1})\left[ x_{1}(u)+q\
x_{2}(u)\right] x_{2}(u)\sinh \eta  \label{lkm.10}
\end{equation}%
where $\beta _{1,2}$, $\beta _{2,1}$ and $\beta _{2,2}$ being the free
parameters and $2\cosh \eta =q+q^{-1}$.

\subsection{The ${\cal U}_{q}[sl(2)]$\ Temperley-Lieb K matrix for $s=1$}

For the biquadratic model \cite{Batch4, KLS}, it follows from (\ref{lkm.4})
and (\ref{lkm.5}) that%
\begin{eqnarray}
K^{-}(u) &=&\left( 
\begin{array}{ccc}
k_{1,1}(u) & \frac{1}{2}\beta _{1,2}\sinh (2u) & \frac{1}{2}\beta
_{1,3}\sinh (2u) \\ 
&  &  \\ 
\frac{1}{2}\beta _{2,1}\sinh (2u) & k_{1,1}(u)+\frac{1}{2}(\beta
_{2,2}-\beta _{1,1})\sinh (2u) & \frac{1}{2}\beta _{2,3}\sinh (2u) \\ 
&  &  \\ 
\frac{1}{2}\beta _{3,1}\sinh (2u) & \frac{1}{2}\beta _{3,2}\sinh (2u) & 
k_{1,1}(u)+\frac{1}{2}(\beta _{3,3}-\beta _{1,1})\sinh (2u)%
\end{array}%
\right)  \nonumber \\
&&  \label{lkm.11}
\end{eqnarray}%
where $k_{1,1}(u)$ is given by (\ref{lkm.6}),%
\begin{eqnarray}
k_{1,1}(u) &=&1-\frac{1}{2}\left\{ (\beta _{3,3}-\beta _{1,1})\left[
x_{1}(u)+q^{2}x_{2}(u)\right] +\frac{\beta _{1,2}\beta _{2,3}}{\beta _{1,3}}%
x_{1}(u)\right.  \nonumber \\
&&\left. +(\beta _{2,2}-\beta _{1,1})x_{2}(u)\right\} x_{2}(u)\sinh \eta .
\label{lkm.12}
\end{eqnarray}%
The diagonal parameters are fixed by the constraint equations (\ref{lkm.7})%
\begin{eqnarray}
\beta _{2,2} &=&\beta _{1,1}+\frac{\beta _{1,2}\beta _{2,3}}{\beta _{13}}-%
\frac{\beta _{21}\beta _{13}}{\beta _{23}},\   \nonumber \\
\beta _{3,3} &=&\beta _{1,1}+\frac{\beta _{1,3}\beta _{3,2}}{\beta _{1,2}}-%
\frac{\beta _{2,1}\beta _{1,3}}{\beta _{2,3}},\   \label{lkm.13}
\end{eqnarray}%
and $\beta _{11}$ is fixed by the normal condition. Moreover, all remained
constraint equations are solved by the relation%
\begin{equation}
\beta _{3,1}=\beta _{3,2}\beta _{2,1}\frac{\beta _{1,3}}{\beta _{1,2}\beta
_{2,3}}.
\end{equation}%
and we have get a five free parameter solution. Here $2\cosh \eta
=q^{-2}+1+q^{2}$ and we made the choice \ $\beta _{1,2}$, $\beta _{1,3}$, $%
\beta _{2,1}$, $\beta _{2,3}$ and $\beta _{3,2}$\ \ for the free parameters.

\subsection{The ${\cal U}_{q}[sl(2)]$ Temperley-Lieb K matrix for $s=\frac{3%
}{2}$}

For $s=\frac{3}{2}$, we have from (\ref{lkm.4}) to (\ref{lkm.6}) the
following non-diagonal entries%
\begin{equation}
k_{i,j}(u)=\frac{1}{2}\beta _{i,j}\sinh (2u),\quad (i\neq j=1,2,3,4)
\label{lkm.14}
\end{equation}%
and the diagonal one%
\begin{equation}
k_{i,i}(u)=k_{1,1}(u)+\frac{1}{2}(\beta _{i,i}-\beta _{1,1})\sinh
(2u),\qquad (i=2,3,4)  \label{lkm.15}
\end{equation}%
with 
\begin{eqnarray}
k_{1,1}(u) &=&1-\frac{1}{2}\left\{ (\beta _{4,4}-\beta _{1,1})\left[
x_{1}(u)+q^{3}x_{2}(u)\right] +\frac{\beta _{1,2}\beta _{2,4}+\beta
_{1,3}\beta _{3,4}}{\beta _{1,4}}x_{1}(u)\right.  \nonumber \\
&&\left. +[(\beta _{2,2}-\beta _{1,1})q^{-1}+(\beta _{3,3}-\beta
_{1,1})q]x_{2}(u)\right\} x_{2}(u)\sinh \eta .  \label{lkm.16}
\end{eqnarray}%
From the block $B[2,3]$ (\ref{lkm.7}), we choose to fix the following
diagonal parameters:%
\begin{eqnarray}
\beta _{2,2} &=&\beta _{1,1}+\frac{\beta _{1,2}\beta _{2,4}+\beta
_{1,3}\beta _{3,4}}{\beta _{1,4}}-\frac{\beta _{2,1}\beta _{1,4}+\beta
_{2,3}\beta _{3,4}}{\beta _{2,4}},  \nonumber \\
\beta _{3,3} &=&\beta _{1,1}+\frac{\beta _{1,2}\beta _{2,4}+\beta
_{1,3}\beta _{3,4}}{\beta _{1,4}}-\frac{\beta _{3,1}\beta _{1,4}+\beta
_{3,2}\beta _{2,4}}{\beta _{3,4}},  \nonumber \\
\beta _{4,4} &=&\beta _{1,1}+\frac{\beta _{1,2}\beta _{2,3}+\beta
_{1,4}\beta _{4,3}}{\beta _{1,3}}-\frac{\beta _{3,1}\beta _{1,4}+\beta
_{3,2}\beta _{2,4}}{\beta _{3,4}}.  \label{lkm.17}
\end{eqnarray}%
All remained constraint equations are solved by the choice%
\begin{eqnarray}
\beta _{4,1} &=&\frac{\beta _{4,2}\beta _{2,1}+\beta _{4,3}\beta _{3,1}}{%
\beta _{1,2}\beta _{2,4}+\beta _{1,3}\beta _{3,4}}\beta _{1,4},  \nonumber \\
\beta _{3,2} &=&-\frac{\beta _{3,1}\beta _{1,2}+\beta _{3,4}\beta _{4,2}}{%
\beta _{1,2}\beta _{2,4}+\beta _{1,3}\beta _{3,4}}\beta _{1,4},  \nonumber \\
\beta _{2,3} &=&-\frac{\beta _{2,1}\beta _{1,3}+\beta _{2,4}\beta _{4,3}}{%
\beta _{1,2}\beta _{2,4}+\beta _{1,3}\beta _{3,4}}\beta _{1,4},
\label{lkm.18}
\end{eqnarray}%
It means that we have found a $K^{-}$ matrix with nine free parameters.

Now we notice that from now on that we will be using the functions of the
type%
\begin{equation}
\Psi _{i,j}=\frac{1}{\beta _{i,j}}\sum_{k\neq i\neq j}^{4}\beta _{i,k}\beta
_{k,j}
\end{equation}%
so that (\ref{lkm.17}) and (\ref{lkm.18}) can be written as 
\begin{eqnarray}
\beta _{2,2} &=&\beta _{1,1}+\Psi _{1,4}-\Psi _{2,4},\quad \beta
_{3,3}=\beta _{1,1}+\Psi _{1,4}-\Psi _{3,4},  \nonumber \\
\beta _{4,4} &=&\beta _{1,1}+\Psi _{1,3}-\Psi _{3,4}
\end{eqnarray}%
and%
\begin{equation}
\Psi _{4,1}=\Psi _{1,4},\quad \Psi _{3,2}=-\Psi _{1,4},\quad \Psi
_{2,3}=-\Psi _{1,4},
\end{equation}%
respectively.

\subsection{The ${\cal U}_{q}[sl(2)]$ Temperley-Lieb K matrix for $s=2$}

\bigskip For $s=2$, the matrix elements are%
\begin{equation}
k_{i,j}(u)=\frac{1}{2}\beta _{i,j}\sinh (2u),\quad (i\neq j=1,...,5)
\label{lkm.19}
\end{equation}%
and%
\begin{equation}
k_{i,i}(u)=k_{1,1}(u)+\frac{1}{2}(\beta _{i,i}-\beta _{1,1})\sinh
(2u),\qquad (i=2,...,5).  \label{lkm.20}
\end{equation}%
where%
\begin{eqnarray}
k_{1,1}(u) &=&1-\frac{1}{2}\left\{ (\beta _{5,5}-\beta _{1,1})\left[
x_{1}(u)+q^{4}x_{2}(u)\right] +\frac{\beta _{1,2}\beta _{2,5}+\beta
_{1,3}\beta _{3,5}+\beta _{1,4}\beta _{4,5}}{\beta _{1,5}}x_{1}(u)\right. 
\nonumber \\
&&\left. +[(\beta _{2,2}-\beta _{1,1})q^{-2}+(\beta _{3,3}-\beta
_{1,1})+(\beta _{4,4}-\beta _{1,1})q^{2}]x_{2}(u)\right\} x_{2}(u)\sinh \eta
\label{lkm.21}
\end{eqnarray}

Substituting these expressions into the reflection equations belonging to
the blocks (\ref{lkm.7}), {\it i.e.,} $B[2,5]$, $B[3,5]$ and $B[4,5]$, we
can find the diagonal parameters 
\begin{eqnarray}
\beta _{2,2} &=&\beta _{1,1}+\Psi _{1,5}-\Psi _{2,5},\qquad \beta
_{3,3}=\beta _{1,1}+\Psi _{1,5}-\Psi _{3,5}  \nonumber \\
\beta _{4,4} &=&\beta _{1,1}+\Psi _{1,5}-\Psi _{4,5},\qquad \beta
_{5,5}=\beta _{1,1}+\Psi _{1,4}-\Psi _{4,5}  \label{lkm.22}
\end{eqnarray}%
where we have defined the twenty functions 
\begin{equation}
\Psi _{i,j}=\frac{1}{\beta _{i,j}}\sum_{k\neq i\neq j}^{5}\beta _{i,k}\beta
_{k,j}  \label{lkm.23}
\end{equation}%
From the blocks $B[i,k]$ $(k\geq i=1,2,3,4)$, we can see that all constraint
equations are rewritten by ten symmetric relations%
\begin{equation}
\Psi _{j,i}=\Psi _{i,j}\qquad (j>i)  \label{lkm.24}
\end{equation}%
and the five relations 
\begin{eqnarray}
\Psi _{2,4} &=&\Psi _{2,3}+\Psi _{1,4}-\Psi _{1,3},\qquad \Psi _{2,5}=\Psi
_{2,3}+\Psi _{1,5}-\Psi _{1,3},  \nonumber \\
\Psi _{3,4} &=&\Psi _{2,3}+\Psi _{1,4}-\Psi _{1,2},\qquad \Psi _{3,5}=\Psi
_{2,3}+\Psi _{1,5}-\Psi _{1,2},  \nonumber \\
\Psi _{4,5} &=&\Psi _{2,3}+\Psi _{1,4}+\Psi _{1,5}-\Psi _{1,2}-\Psi _{1,3}.
\label{lkm.25}
\end{eqnarray}

Now, all we need is to look for the constraint equations belonging to the
blocks $B[5,k]$. There are seven remained equations but only four are
independent%
\begin{eqnarray}
\Theta _{2,2} &=&\Theta _{1,1}-\left( \Psi _{2,3}-\Psi _{1,3}\right) \Psi
_{1,2}  \nonumber \\
\Theta _{3,3} &=&\Theta _{1,1}-\left( \Psi _{2,3}-\Psi _{1,2}\right) \Psi
_{1,3}  \nonumber \\
\Theta _{4,4} &=&\Theta _{1,1}-\left( \Psi _{2,3}+\Psi _{1,4}-\Psi
_{1,2}-\Psi _{1,3}\right) \Psi _{1,4}  \nonumber \\
\Theta _{5,5} &=&\Theta _{1,1}-\left( \Psi _{2,3}+\Psi _{1,5}-\Psi
_{1,2}-\Psi _{1,3}\right) \Psi _{1,5}  \label{lkm.27}
\end{eqnarray}%
where we have defined five new functions of the type%
\begin{equation}
\Theta _{i,i}=\sum_{k\neq i}^{5}\beta _{i,k}\beta _{k,i},\qquad i=1,...,5
\label{lkm.28}
\end{equation}%
Finally, all constraint equations are substituted by these relations.

Substituting (\ref{lkm.25}) into the ten symmetric relations (\ref{lkm.24})
we still have to solve them in order to write explicitly the $K^{-}$
solution. A way to do this is to solve simultaneously these ten recursive
relations and take into account the solution with the smallest numbers of
fixed parameters. Following this procedure we have find a unique solutions
with seven fixed parameters. for instance, $\beta _{1,2}$, $\beta _{1,5}$, $%
\beta _{2,4}$, $\beta _{2,5}$, $\beta _{3,4}$, $\beta _{3.5}$, and $\beta
_{4,5}$. Its expressions in terms of the free parameters are not written
here because they are too large and cumbersome. These seven parameters plus
the four diagonal parameters (\ref{lkm.22}) and the normal condition, give
us a $5$ by $5$ reflection $K^{-}$ matrix solution with $13$ free parameters!

\subsection{The ${\cal U}_{q}[sl(2)]$\ Temperley-Lieb K-matrix for $s>2$}

The main difficulty is to solve the constraint equations, which the solution
will to find some of the parameters in the solution given by (\ref{lkm.4})-(%
\ref{lkm.6}). The existence of many free parameters turn out that it is a
very hard task. The constraint equations are big and became biggest after
any algebraic manipulation so that our computer resources have became weak
or not enough sufficient in order to work with too large algebraic
expressions. In its original form, it is a very formidable problem.

In order to proceed with this task we have introduced two new objects
instead of working directly with the $\beta _{i,j}$ parameters 
\begin{equation}
\Psi _{i,j}=\frac{1}{\beta _{i,j}}\sum_{k\neq i,j}\beta _{i,k}\beta
_{k,j}\qquad {\rm and}\qquad \Theta _{i,i}=\sum_{k\neq i}\beta _{i,k}\beta
_{k,i}\text{.}  \label{spins.1}
\end{equation}%
After we rewrite the constraint equations in terms of $\Psi _{i,j}$ and $%
\Theta _{i,i}$, we can easily solve the blocks (\ref{lkm.7}) in order to
find $2s$ diagonal parameters,%
\begin{eqnarray}
\beta _{i,i} &=&\beta _{1,1}+\Psi _{1,2s+1}-\Psi _{i,2s+1},\qquad
i=2,3,...,2s  \nonumber \\
\beta _{2s+1,2s+1} &=&\beta _{1,1}+\Psi _{1,2s}-\Psi _{2s,2s+1}.
\label{spins.2}
\end{eqnarray}

All equations from the block $B[1,k]$ to the block $B[2s,k]$ are now
substituted by $s(2s+1)$ symmetric relations%
\begin{equation}
\Psi _{j,i}=\Psi _{i,j},\quad j>i  \label{spins.3}
\end{equation}%
and$\ 4(s-1)$ relations of the type%
\begin{eqnarray}
\Psi _{2,j} &=&\Psi _{2,3}+\Psi _{1,j}-\Psi _{1,3},\qquad j=4,...,2s+1, 
\nonumber \\
\Psi _{3,j} &=&\Psi _{2,3}+\Psi _{1,j}-\Psi _{1,2,}\qquad j=4,...,2s+1.
\label{spins.4}
\end{eqnarray}%
The remained equations contained in the block $B[2s+1,k]$ are rewritten by $%
2(s-1)(s-\frac{3}{2})$ relations involving the $\Psi _{i,j}$ 
\begin{equation}
\Psi _{i,j}=\Psi _{1,i}+\Psi _{1,j}-\Psi _{1,2}-\Psi _{1,3},\qquad
i=4,...,2s,\quad j=i+1,...,2s+1  \label{spins.5}
\end{equation}%
and $4s-1$ relations involving the diagonal $\beta _{k,k}$ parameters, $\Psi
_{1,2s+1}$ and the $\Theta _{j,j}$, 
\begin{eqnarray}
\Theta _{j,j} &=&\Theta _{2s+1,2s+1}+(\beta _{2s+1,2s+1}-\beta _{j,j})(\beta
_{j,j}-\beta _{1,1}-\Psi _{1,2s+1}),\quad j=2,3,...,2s,  \nonumber \\
\Theta _{j^{\prime },j^{\prime }} &=&\Theta _{1,1}+(\beta _{1,1}-\beta
_{j^{\prime },j^{\prime }})(\beta _{j^{\prime },j^{\prime }}-\beta
_{2s+1,2s+1}-\Psi _{1,2s+1}),\quad j=2,3,...,2s,  \nonumber \\
\Theta _{2s+1,2s+1} &=&\Theta _{1,1}-(\beta _{1,1}-\beta _{2s+1,2s+1})\Psi
_{1,2s+1},  \label{spins.6}
\end{eqnarray}%
where $j^{\prime }=2s+2-j$.

From (\ref{spins.3}) to (\ref{spins.6}) one can account $4s(s+1)-2$
constraint equations but, after we have substituted the relations (\ref%
{spins.4}) and (\ref{spins.5}) into (\ref{spins.3}), we only need to look
for these $s(2s+1)$ symmetric relations.

From these relations we have fixed $2s^{2}-1$ parameters when $s$ is integer
and $2s^{2}-3/2$ parameters when $s$ is semi-integer.

Again, we recall that the last step requires large computer resources and
the final expressions for the $\beta _{i,j}$ parameters are too large and
cumbersome to be written in any way. \ It means that we have found the
normal reflection matrices for the ${\cal U}_{q}[sl(2)]$ Temperley-Lieb
model. These matrices are solutions of (\ref{int.2}) with $2s(s+1)+1$ free
parameters if $s=1,2,...$ and $2s(s+1)+3/2$ free parameters if $%
s=1/2,3/2,... $ .

\section{Reduced Solutions}

Now, for particular choices of the free parameters, we can derive several
subclasses of solutions from the general solution presented in the previous
section. However, using any reduced procedure we can louse some of the
possible reductions. For instance, it is simpler to solve directly the
reflection equations by looking for diagonal solutions instead of deriving
them from the general one. In the sequence, we present some particular
solutions which can be obtained from the general solution. Let us start with
the diagonal solutions.

\subsection{Diagonal Solutions}

Taking into account only the diagonal $K^{-}$ matrices, the reflection
equations are solved when we find all matrix elements $k_{j,j}(u)$, $%
j=2,...,2s+1$ as function of $k_{1,1}(u)$, provided that the diagonal
parameters $\beta _{j,j}$ satisfy $s(2s-1)$ constraint equations of the type%
\begin{equation}
\left( \beta _{2s+1,2s+1}-\beta _{i,i}\right) \left( \beta
_{2s+1,2s+1}-\beta _{j,j}\right) \left( \beta _{j,j}-\beta _{i,i}\right)
=0\qquad (i\neq j\neq 2s+1)  \label{red.1}
\end{equation}%
From (\ref{red.1}) we can find diagonal $K^{-}$ matrix solutions with only
two type of entries. Let us normalize one of them to be equal to $1$ such
that the other entry is given by%
\begin{equation}
k_{p,p}(u)=-\frac{\beta _{p,p}x_{2}(u)\left[ \Delta _{1}x_{2}(u)+x_{1}(u)%
\right] +2\left[ x_{1}(u)x_{2}^{\prime }(u)-x_{1}^{\prime }(u)x_{2}(u)\right]
}{\beta _{p,p}x_{2}(u)\left[ \Delta _{k}x_{2}(u)+x_{1}(u)\right] -2\left[
x_{1}(u)x_{2}^{\prime }(u)-x_{1}^{\prime }(u)x_{2}(u)\right] }
\label{red.1a}
\end{equation}%
where $\Delta _{1}+\Delta _{k}=2\cosh \eta $.

Identifying the diagonal positions with the powers of $q$, $%
(1,2,...,2s,2s+1)\circeq (q^{-2s},q^{-2s+2},...,q^{2s-2},q^{2s})$ one can
see that $\Delta _{1}$ is the sum of the powers of $q$ corresponding to the
positions of the entries $1$ and $\Delta _{k}$ is the sum of the power of $q$
corresponding to the positions of the entries $k_{p,p}(u)$.

Denoting the diagonal solutions by ${\Bbb K}_{{\bf a}_{s}}^{[r]}$ where $%
{\bf a}_{s}=(a_{1},a_{2},...,a_{2s+1})$ with $a_{i}=0$ if $k_{i,i}(u)=1$ or $%
a_{i}=1$ if $k_{i,i}(u)=k_{pp}(u)$ and $r$ is the number of the entries $%
k_{p,p}(u)$ distributed on diagonal positions and $p$ being the first
position with the entry different from $1$. Thus, we have counted

\begin{equation}
N=\sum\limits_{r=1}^{2s}\frac{(2s+1)!}{r!\left( 2s+1-r\right) !}
\label{red.2}
\end{equation}
for the number of diagonal $K^{-}$ matrix solutions with one free parameter.
Again, the $K^{+}$ solutions are obtained by the isomorphism (\ref{int.4})

Let us work explicitly with this description. For the case $s=\frac{1}{2}$
one can find two diagonal solutions:%
\begin{equation}
{\Bbb K}_{(1,0)}^{[1]}=\left( 
\begin{array}{cc}
k_{1,1}(u) & 0 \\ 
0 & 1%
\end{array}%
\right) ,\qquad {\Bbb K}_{(0,1)}^{[1]}=\left( 
\begin{array}{cc}
1 & 0 \\ 
0 & k_{2,2}(u)%
\end{array}%
\right)  \label{red.3}
\end{equation}%
where%
\begin{eqnarray}
k_{1,1}(u) &=&-\frac{\beta _{1,1}x_{2}(u)\left[ qx_{2}(u)+x_{1}(u)\right] +2%
\left[ x_{1}(u)x_{2}^{\prime }(u)-x_{1}^{\prime }(u)x_{2}(u)\right] }{\beta
_{1,1}x_{2}(u)\left[ q^{-1}x_{2}(u)+x_{1}(u)\right] -2\left[
x_{1}(u)x_{2}^{\prime }(u)-x_{1}^{\prime }(u)x_{2}(u)\right] }  \nonumber \\
k_{2,2}(u) &=&-\frac{\beta _{2,2}x_{2}(u)\left[ q^{-1}x_{2}(u)+x_{1}(u)%
\right] +2\left[ x_{1}(u)x_{2}^{\prime }(u)-x_{1}^{\prime }(u)x_{2}(u)\right]
}{\beta _{2,2}x_{2}(u)\left[ qx_{2}(u)+x_{1}(u)\right] -2\left[
x_{1}(u)x_{2}^{\prime }(u)-x_{1}^{\prime }(u)x_{2}(u)\right] }  \label{red.4}
\end{eqnarray}%
with $2\cosh \eta =q^{-1}+q$. Of course, they are equivalent by the exchange 
$q\leftrightarrow q^{-1}$.

Similarly, for $s=1$, one can find six diagonal solutions, half of them with
one entry different from $1$%
\begin{eqnarray}
{\Bbb K}_{(1,0,0)}^{[1]} &=&\left( 
\begin{array}{ccc}
k_{1,1}(u) & 0 & 0 \\ 
0 & 1 & 0 \\ 
0 & 0 & 1%
\end{array}%
\right) ,\quad {\Bbb K}_{(0,1,0)}^{[1]}=\left( 
\begin{array}{ccc}
1 & 0 & 0 \\ 
0 & k_{2,2}(u) & 0 \\ 
0 & 0 & 1%
\end{array}%
\right) ,\quad  \nonumber \\
{\Bbb K}_{(0,0,1)}^{[1]} &=&\left( 
\begin{array}{ccc}
1 & 0 & 0 \\ 
0 & 1 & 0 \\ 
0 & 0 & k_{3,3}(u)%
\end{array}%
\right)  \label{red.5}
\end{eqnarray}%
with%
\begin{eqnarray}
k_{1,1}(u) &=&-\frac{\beta _{1,1}x_{2}(u)\left[ (1+q^{2})x_{2}(u)+x_{1}(u)%
\right] +2\left[ x_{1}(u)x_{2}^{\prime }(u)-x_{1}^{\prime }(u)x_{2}(u)\right]
}{\beta _{1,1}x_{2}(u)\left[ q^{-2}x_{2}(u)+x_{1}(u)\right] -2\left[
x_{1}(u)x_{2}^{\prime }(u)-x_{1}^{\prime }(u)x_{2}(u)\right] }  \nonumber \\
k_{2,2}(u) &=&-\frac{\beta _{2,2}x_{2}(u)\left[
(q^{-2}+q^{2})x_{2}(u)+x_{1}(u)\right] +2\left[ x_{1}(u)x_{2}^{\prime
}(u)-x_{1}^{\prime }(u)x_{2}(u)\right] }{\beta _{2,2}x_{2}(u)\left[
x_{2}(u)+x_{1}(u)\right] -2\left[ x_{1}(u)x_{2}^{\prime }(u)-x_{1}^{\prime
}(u)x_{2}(u)\right] }  \nonumber \\
k_{3,3}(u) &=&-\frac{\beta _{3,3}x_{2}(u)\left[ (q^{-2}+1)x_{2}(u)+x_{1}(u)%
\right] +2\left[ x_{1}(u)x_{2}^{\prime }(u)-x_{1}^{\prime }(u)x_{2}(u)\right]
}{\beta _{3,3}x_{2}(u)\left[ q^{2}x_{2}(u)+x_{1}(u)\right] -2\left[
x_{1}(u)x_{2}^{\prime }(u)-x_{1}^{\prime }(u)x_{2}(u)\right] }  \label{red.6}
\end{eqnarray}%
and three further diagonal solutions with two equal entries different from
unity%
\begin{eqnarray}
{\Bbb K}_{(1,1,0)}^{[2]} &=&\left( 
\begin{array}{ccc}
k_{1,1}(u) & 0 & 0 \\ 
0 & k_{1,1}(u) & 0 \\ 
0 & 0 & 1%
\end{array}%
\right) ,\quad {\Bbb K}_{(1,0,1)}^{[2]}=\left( 
\begin{array}{ccc}
{\bf k}_{1,1}(u) & 0 & 0 \\ 
0 & 1 & 0 \\ 
0 & 0 & {\bf k}_{1,1}(u)%
\end{array}%
\right) ,  \nonumber \\
{\Bbb K}_{(0,1,1)}^{[2]} &=&\left( 
\begin{array}{ccc}
1 & 0 & 0 \\ 
0 & k_{2,2}(u) & 0 \\ 
0 & 0 & k_{2,2}(u)%
\end{array}%
\right)  \label{red.7}
\end{eqnarray}%
where%
\begin{eqnarray}
k_{1,1}(u) &=&-\frac{\beta _{1,1}x_{2}(u)\left[ q^{2}x_{2}(u)+x_{1}(u)\right]
+2\left[ x_{1}(u)x_{2}^{\prime }(u)-x_{1}^{\prime }(u)x_{2}(u)\right] }{%
\beta _{1,1}x_{2}(u)\left[ (q^{-2}+1)x_{2}(u)+x_{1}(u)\right] -2\left[
x_{1}(u)x_{2}^{\prime }(u)-x_{1}^{\prime }(u)x_{2}(u)\right] },  \nonumber \\
{\bf k}_{1,1}(u) &=&-\frac{{\bf \beta }_{1,1}x_{2}(u)\left[ x_{2}(u)+x_{1}(u)%
\right] +2\left[ x_{1}(u)x_{2}^{\prime }(u)-x_{1}^{\prime }(u)x_{2}(u)\right]
}{{\bf \beta }_{1,1}x_{2}(u)\left[ (q^{-2}+q^{2})x_{2}(u)+x_{1}(u)\right] -2%
\left[ x_{1}(u)x_{2}^{\prime }(u)-x_{1}^{\prime }(u)x_{2}(u)\right] }, 
\nonumber \\
k_{2,2}(u) &=&-\frac{\beta _{2,2}x_{2}(u)\left[ q^{-2}x_{2}(u)+x_{1}(u)%
\right] +2\left[ x_{1}(u)x_{2}^{\prime }(u)-x_{1}^{\prime }(u)x_{2}(u)\right]
}{\beta _{2,2}x_{2}(u)\left[ (1+q^{2})x_{2}(u)+x_{1}(u)\right] -2\left[
x_{1}(u)x_{2}^{\prime }(u)-x_{1}^{\prime }(u)x_{2}(u)\right] }.
\label{red.8}
\end{eqnarray}%
Here we notice again that the difference between the entries (\ref{red.8})
come from the partitions of $2\cosh \eta $ such that $\Delta _{1}+\Delta
_{k}=q^{-2}+1+q^{2}$ and the equivalence due to the symmetry $%
q\leftrightarrow q^{-1}$. It follows on and on similarly for the other
values of $s$.

\subsection{Z$_{2s+1}$ K matrix solution}

Now we recall the center of ${\cal U}_{q}[sl(2)]$ in order to take into
account the $K^{-}$ matrix solution in terms of the generators of the $%
Z_{2s+1}-$symmetry. \ From the general solution it is straightforward to get
the following reduced solution 
\begin{equation}
K^{-}(u)=k_{1,1}(u)I+\sum_{k=1}^{2s}k_{1,2s+1}(u)\omega ^{k}Z_{k}
\label{red.9}
\end{equation}%
where $I$ is the unity matrix , $\omega =\exp (\frac{2i\pi }{2s+1})$ and $%
Z_{k}$ are the $Z_{2s+1}$ matrices%
\begin{equation}
\left( Z_{k}\right) _{i,j}=\delta _{i,j+k}\qquad \mod(2s+1),\qquad
k=1,...,2s  \label{red.9a}
\end{equation}%
and%
\begin{equation}
k_{1,1}(u)=\frac{k_{1,2s+1}(u)}{\beta _{1,2s+1}[x_{2}(u)\cosh \eta +x_{1}(u)]%
}\left\{ \frac{x_{1}(u)x_{2}^{\prime }(x)-x_{1}^{\prime }(u)x_{2}(u)}{%
x_{2}(u)}-\frac{(2s-1)\omega ^{2s}}{2}x_{1}(u)\right\} .  \label{red.9c}
\end{equation}%
Here $\beta _{1,2s+1}$ is the free parameter and $k_{1,2s+1}(u)$ is an
arbitrary function.

\subsection{Spin zero K matrix solution}

\bigskip Let us to start again with the $2s+1$ by $2s+1$ Temperley-Lieb $R$
matrix (\ref{mod.1}). However, looking for the $K^{-}$ matrix solutions with
the form%
\begin{equation}
K^{-}(u)=f_{1}(u)I^{\prime }+f_{2}(u)U^{\prime }  \label{red.11}
\end{equation}%
where $f_{1}(u)$ and $f_{2}(u)$ are functions to be determined in order that 
$U^{\prime }$ is the spin zero projector (\ref{mod.3}) for the spin $%
s^{\prime }$ and $I^{\prime }$ being the corresponding identity matrix.

When $(2s^{\prime }+1)^{2}=2s+1$, we have found that (\ref{red.11}) is
solutions of the reflection equation (\ref{int.2}) provided that

\begin{eqnarray}
f_{1}(u) &=&\frac{f_{2}(u)}{\beta \lbrack x_{2}(u)\cosh \eta +x_{1}(u)]}%
\left\{ \frac{x_{1}(u)x_{2}^{\prime }(x)-x_{1}^{\prime }(u)x_{2}(u)}{x_{2}(u)%
}-\frac{\beta }{2}\sum_{j=-s^{\prime }}^{s^{\prime }}\left[ q^{2j(2s^{\prime
}+1)}x_{2}(u)+q^{2j}x_{1}(u)\right] \right\}  \nonumber \\
&&  \label{red.12}
\end{eqnarray}%
where $\beta =\frac{df_{2}(u)}{du}|_{u=0}$ is the free parameter and $%
f_{2}(u)$ is an arbitrary function.

\section{Conclusion}

In this work we have presented the general solutions of the reflection
equation for the ${\cal U}_{q}[sl(2)]$ Temperley-Lieb vertex model. Our
findings can be summarized into two classes of general solutions depending
if the spin $s$ is a number integer or semi-integer. The large number of
free parameters, $2s(s+1)+1$ for $s$ integer and $2s(s+1)+3/2$ for $s$
semi-integer, is an important property of this model which follows due to
the explicit factorization of the constraint $\delta (2$cosh$\eta -[2s+1])$
in the reflection equation.

These results pave the way to construct, solve and study physical properties
of the underlying quantum spin chains with open boundaries, generalizing the
previous efforts made for the case of periodic boundary conditions \cite%
{KLS2, LS2}. Although we do not know the Algebraic Bethe Ansatz for $s\neq 
\frac{1}{2}$, even for the periodic cases, we expect that the coordinate
Bethe ansatz solution of the Temperley-Lieb models constructed from diagonal
solutions presented here can be obtained by adapting the results of \cite%
{LSG} and the algebraic-funcional method presented in \cite{Galleas} may be
a possibility to treat the non-diagonal cases. We expect the results
presented here to motivate further developments on the subject of integrable
open boundaries for the Temperley-Lieb vertex models based on other $q$%
-deformed Lie algebras \cite{Batch3} and superalgebras \cite{Zhang}.

\vspace{.35cm}%

{\bf Acknowledgment:} This work was supported in part by Funda\c{c}\~{a}o de
Amparo \`{a} Pesquisa do Estado de S\~{a}o Paulo-{\small FAPESP}-Brasil and
by Conselho Nacional de Desenvolvimento-{\small CNPq}-Brasil.

\end{document}